\documentclass[preprint,12pt]{elsarticle}
\usepackage{amssymb}
\usepackage{inputenc}
\usepackage{url}
\usepackage{array}
\usepackage{graphicx}
\usepackage{pstricks}
\usepackage{pst-node}
\usepackage{amsmath}
\usepackage{eurosym}
\usepackage{times}
\usepackage{hyperref}
\usepackage{color}
\usepackage[T1]{fontenc}
\usepackage{helvet}

\usepackage{subfigure}

\usepackage{latexsym}

\journal{Physica A}

\begin{document}

\begin{frontmatter}



\title{Generalization of the Aoki-Yoshikawa sectoral productivity model based on extreme physical information principle\tnoteref{foot}
}
\tnotetext[foot]{Highlights:\\
Assumptions of original Aoki-Yoshikawa sectoral productivity model (AYM) are given. \\
Information channel capacity for the AYM in the extremal physical information (EPI) method is constructed.\\
Analytical observed structural principle and variational one are found.\\
Generating equation for AYM and probability distribution for AYM are found.\\
Results of the original AYM and of the EPI method approach to AYM are compared.
}
\author{Ilona Bednarek${}^{\, a}$}
\ead{ilona.bednarek@us.edu.pl}
\author{
Marcin Makowski${}^{\,b}$}
\ead{makowski.m@gmail.com}
\author{
Edward W. Piotrowski${}^{\,b}$}
\ead{qmgames@gmail.com}
\author{
        Jan S{\l}adkowski${}^{\,a}$
\corref{cor1}\fnref{label1}
}
\ead{jan.sladkowski@us.edu.pl}
\author{
Jacek Syska${}^{a}$
\corref{cor2}\fnref{label2}
}
\ead{jacek.syska@us.edu.pl}

\fntext[label1]{corresponding author}
\fntext[label2]{corresponding author}

\fntext[label3]{Published in: Physica A 428 (2015) 161-172. http://dx.doi.org/10.1016/j.physa.2015.02.033}

\address{${}^{\,a }$ Institute of Physics, University of Silesia, Uniwersytecka 4, Pl 40-007 Katowice, Poland}

\address{${}^{\,b}$ Institute of Mathematics, University of Bialystok, Bialystok, Poland}



\begin{abstract}
This paper presents a continuous variable generalization of the Aoki-Yoshikawa sectoral productivity model. Information theoretical methods from the Frieden-Soffer extreme physical information  statistical estimation methodology were used to construct exact solutions. Both approaches coincide in first order approximation. The approach proposed here can be successfully applied in other fields of research.
\end{abstract}

\end{frontmatter}


\newpage

\section{Introduction}

\label{Introduction}

Much of economic theory is currently discussed  in terms of mathematical economic models. Mathematical economics aims at representation and analysis of problems in economics in order to form meaningful and testable propositions about complex issues often described in a less formal way in everyday life. Econophysics, on the other hand, originates from attempts at solving problems in economics with tools developed by physicists, and is evolving into an interdisciplinary research field. Several applications of the approach to stylized models of economics have been recently put forward \cite{garibaldi,aoy}. The aim of this paper is to show how the extremal physical information (EPI) method of Frieden and Soffer \cite{Frieden-Soffer,frieden} can be used to develop a generalization of the Aoki-Yoshikawa  sectoral productivity model (AYM) \cite{aoki,garibaldi}. Below, its modified \cite{Dziekuje informacja_2,Dziekuje informacja_1} version, which abandons the previous, arbitrary metrical form  is presented and the solution to entailed equations of the fully analytical formulation of the information principles problem \cite{Dziekuje informacja_2} is given. The approach is based on the maximum-likelihood estimation (MLE) and the Fisher information, both the observed and expected ones, defined as the expectation value of the observed information widely used in information geometry \cite{Amari Nagaoka book} and  statistics \cite{frieden,Dziekuje informacja_1,pss,Dziekuje informacja_2}. A similar approach was previously used to analyse the problem of subjectivity in supply-demand related issues \cite{ps,pss}.
This paper is organized as follows.  In Section~\ref{ayspm}  the  original formulation of the AYM
 sectoral productivity model is presented. In
Section~\ref{epim}  generalization of the method based on the
EPI method is introduced. In Section~\ref{porownanie} both approaches are compared and, finally, in Section~\ref{Conclusions} conclusions are drawn.

\section{The Aoki-Yoshikawa Sectoral Productivity Model}

\label{ayspm}

Sectoral productivity models  form  key issues  in  the analysis of productivity growth at an
intermediate level of aggregation \cite{aoki}. Such analyses aim to describe the patterns of productivity growth
across and within sectors (e.g. agriculture, manufacturing, and services)
and to identify main policy factors driving these patterns. The natural way of presenting
these models is in terms of transition probabilities over occupation states.
In such an approach, occupation vectors and partition vectors can be given
an interpretation in terms of economic variables. In a mathematical approach, irreducible
and aperiodic Markov chains that have a unique invariant distribution
are used \cite{garibaldi,Scalas-Garibaldi-2}. Such an approach makes it possible to cope with a large number of interacting heterogeneous agents and to some extent ignores the issue of rationality of  agents' behaviour as it is impossible to follow the ``motion''
of an individual agent in a system composed of about $10^6$ individuals. Therefore, the assumption that precise behaviour of each agent is irrelevant, is the crucial point. This enables one to adopt some techniques used in statistical physics and it follows that some models of macroeconomics can be built on analogous premises. In their  book \cite{aoki}, Masanao Aoki and Hiroshi Yoshikawa presented, among others, an interesting model for the economy of a country with $g$ economic sectors. The $i$th sector is characterized by the amount of production factor $n_i$, that is, the number of workers in the sector $i$,  and by its level of productivity (effectiveness) $a_i$.

Aoki and Yoshikawa were interested in finding
the probability distribution of the productivity among sectors.
In the statistical physics language this means that the
probability distribution of the occupation vector
 \begin{equation}
\label{rozklad n}
\vec{n} = (n_{1} , n_{2} , ... , n_{g}) \;
\end{equation}
of the system is searched.
This coincides with the standard statistical physics problem of finding allocations of $n$ particles to $g$ energy levels.
According to Boltzmann,
the probability
distribution of the occupation vector is equal to
\cite{aoki}
\begin{eqnarray}
\label{prawdopodobienstwo pojawieniem sie wektora n}
\pi({\vec n}) =  \frac{n!}{\prod_{i=1}^{g} n_{i}!} \prod_{i=1}^{g} p^{n_i}
= \frac{n!}{\prod_{i=1}^{g} n_{i}!} \;p^{n}
\; ,
\end{eqnarray}
where $p$ is the probability of the occupation of the particular $s$th sector ($s = 1, 2, ..., g$) by the $i$th worker taken to be the same for all particular configurations of these occupations.

The total production factor is exogenously given and fixed so that
  \begin{eqnarray}
\label{calkowita produkcji}
\sum_{i=1}^{g} n_{i} = n \; , \ \ \  (n \; \texttt{{fixed}}).
\end{eqnarray}
The output of the $i$-th sector is given by
\begin{eqnarray}
\label{produkcja w i tym sektorze}
z_{i} = a_{i} \, n_{i}. \;
\end{eqnarray}
The Gross Domestic Product (GPD), that is, the total output of the country $Z$ is equal to
\begin{eqnarray}
\label{calkowity uzysk z}
Z = \sum_{i=1}^{g} z_{i} = \sum_{i=1}^{g} a_{i} \, n_{i} \; .
\end{eqnarray}
In the model it is equal to exogenous aggregated demand $D$, i.e.:
\begin{eqnarray}
\label{Z rowne D}
Z = D  \; , \ \ \  (D \; \texttt{{fixed}}).
\end{eqnarray}
{\bf Note.} It is possible to consider versions of AYM where the demand bound (\ref{Z rowne D}) or the condition of constancy of number of workers (\ref{calkowita produkcji}) in the system are relaxed \cite{Scalas-Garibaldi-2}.
%

The standard Lagrange multipliers method can be used to find the occupation vector
${\vec n}$ which maximizes the probability $\pi({\vec n})$
with conserved both the total production factor $n$ and the total GPD equal to $D$.
With the help of the  Stirling for\-mula $\ln ({\prod_{i=1}^{g} n_{i}!} )$ $= \sum _{i=1}^{g} n_{i}(\ln n_{i} -1)\ \ { (n>>1\ \texttt{
and}\  \ n_{i}>>1 )}$ the problem is reduced to finding  the solution of the system of $g$ equations:
\begin{eqnarray}
\label{MNW z warunkami}
\frac{\partial  }{\partial n_{i}} \left[ \ln \pi({\vec n})  + \nu \left( \sum_{i=1}^{g}n_{i} -n \right) - \beta \left( \sum_{i=1}^{g} a_{i}n_{i} - D \right) \right] = 0 \; .
\end{eqnarray}
The solution has the following form:
 \begin{eqnarray}
\label{n*}
n_{i} = n_{i}^{*} = e^{\nu} \, e^{-\beta a_{i}} \; , \;\;\; i=1,2,...,g \; .
\end{eqnarray}
The constants $\nu$ and $\beta$ are determined by inserting (\ref{n*}) in (\ref{calkowita produkcji}) and (\ref{calkowity uzysk z})-(\ref{Z rowne D}).
This is the Boltzmann distribution for the system which is in the state of the statistical equilibrium.
Scalas and Garibaldi showed \cite{Scalas-Garibaldi-2} that
there is a more general solution of the form
\begin{eqnarray}
\label{n**}
n_{i} = n_{i}^{**} = \frac{1}{e^{-\nu} \, e^{\beta a_{i}}-c }\; , \;\; i=1,2,...,g; \; c\in \mathbb{R} \, ,
\end{eqnarray}
where $c$ is a parameter. Eq.(\ref{n**}) arises when the appropriate Markovian dynamics is taken into account,
with the transition probabilities which are tuning, via their dependance on the parameter $c$, the choice of a new productivity sector for workers leaving their sector \cite{Scalas-Garibaldi-2}.
%
%
Only in case of $c = 0$ the Aoki and Yoshikawa solution (\ref{n*}) is recovered and the interpretation of the cases when
$c \neq 0$ can be found in \cite{Scalas-Garibaldi-2}.

With the further assumption \cite{garibaldi}:
\begin{eqnarray}
\label{ciag a i}
a_{i} = i \; a_{0} \; \;\;\;  \;\;\; i=1,2,...,g \;,
\end{eqnarray}
where $a_{0}$ denotes the minimal productivity, one gets the
most probable vector
\begin{eqnarray}
\label{wynik AY na p}
n_{i}^{*} =  \frac{n}{r-1} \left( \frac{r-1}{r} \right)^{i}\; ,
\end{eqnarray}
where $r = \frac{D/n}{a_{0}}$ is the
aggregated demand $D$ per agent divided by the minimal productivity.
In the limit $r\gg 1$ one gets\begin{eqnarray}
\label{wynik AY2 na p2}
n_{i}^{*} =  \frac{n}{r-1} \left( \frac{r-1}{r} \right)^{i}\approx (\frac{1}{r} + \frac{1}{r^2}) \; e^{ -  \frac{i}{r}  } \; , \;\;\; i=1,2,... \;  ;  \;\;\; r \gg 1 \; .
\end{eqnarray}
The assumption $a_{i} = i\, a_{0} $ also allows for a simplified worker dynamics via creation and annihilation of components of the occupation vector but due care must be taken to make sure that $n$ is conserved \cite{Scalas-Garibaldi-2}. 
Alternatively, in the EPI method approach to the AYM model, the probability distribution
$p\left( a \right)$ of the level $a$ of productivity will be found
(see Eq.(\ref{rozklad koncowy A})).
If the probability distribution
$p\left(t, a \right)$, which is normalized over space and time
$\int_{0}^{T} \int_{{\cal Y}_{a}} dt \, da \, p\left(t, a \right) =1$, then
$p\left(t \right)  = \int_{{\cal Y}_{a}} \, da \, p\left(t, a \right)$
represents the probability
that the worker is found at the time $(t, t+dt)$ somewhere within measurement productivity space ${\cal Y}_{a}$, which is a set of possible values of productivity.
For example, a high $p(t)\, dt$ means that there is a high chance that a worker is found
anywhere in the space of productivities at this time.
In particle physics this property could be called
probabilistic creation \cite{frieden}.
%
%
%

\section[Basic information on the information channel capacity]{Basic information on the information channel capacity}

\label{basic FI}

Suppose that the original random variable $Y$ takes
vector values ${\bf y} \in {\cal Y}$ and let the
$k$ - dimensional vector parameter $\theta$ of
the distribution $p({\bf y})$
be the {\it expected parameter}, i.e., the expectation value of $Y$:
\begin{eqnarray}
\label{wartosc oczekiwana EY}
\theta \equiv E(Y)= \int_{\cal Y} d {\bf y}\, p({\bf y})\,  {\bf y} \; .
\end{eqnarray}
Let us now consider the $N$-dimensional sample $\widetilde{Y} = (Y_{1},Y_{2},...,Y_{N}) \equiv( Y_{n})_{n=1}^{N}$, where every $Y_{n}$ is the variable $Y$ in the $n$th population, $n=1,2,...,N$, which is characterized by the value of the vector parameter $\theta_{n}$.
The specific realization of $\widetilde{Y}$ takes the form
$y=({{\bf y}_{1},{\bf y}_{2},...,{\bf y}_{N}})\equiv ({\bf y}_{n})_{n=1}^{N}$, where
every datum ${\bf y}_{n}$ is generated from the distribution $p_{n}({\bf y}_{n}|\Theta)$
of the random variable $Y_{n}$, where
the $d=k \times N$ - dimensional vector parameter  $\Theta$ \cite{Dziekuje za channel} is given by:
\begin{eqnarray}
\label{parametr Theta}
\Theta = (\theta_{1}, \theta_{2},...,\theta_{N})^{T} \equiv (\theta_{n})_{n=1}^{N} \; .
\end{eqnarray}
The set of all possible realizations $y$ of the sample $\widetilde{Y}$ forms the sample space ${\cal B}$ of the system.
%
%
When the variables $Y_{n}$ of the sample $\widetilde{Y}$ are independent, then the expected parameter
$\theta_{n'}  =  \int_{\cal B} dy \, P(y|\Theta) \, {\bf y}_{n'} $ does not influence  the {\it point probability} distribution $p_{n}({\bf y}_{n}|\theta_{n})$ for the sample index $n' \neq n$.
The data are generated in agreement with the point probability distributions, which fulfill the condition:
\begin{eqnarray}
\label{rozklady punktowe polozenia}
p_{n}({\bf y}_{n}|\Theta) = p_{n}({\bf y}_{n}|{\theta}_{n}) \; , \;\;\;\; {\rm where} \;\;\; n=1,...,N \; ,
\end{eqnarray}
and {\it the likelihood function} $P(y\,|\Theta)$ of the sample $y = ({\bf y}_{n})_{n=1}^{N}$ is the product:
\begin{eqnarray}
\label{funkcja wiarygodnosci proby - def}
P(\Theta) \equiv P\left({y|\Theta}\right) = \prod\limits_{n=1}^{N} {p_{n}\left({{\bf y}_{n}|\theta_{n}}\right)} \; .
\end{eqnarray}

{\bf The Fisher information matrix}:
%
%
Assume that on ${\cal B}$ the $d$~-~dimensional statistical model:
\begin{eqnarray}
\label{model statystyczny S}
{\cal S} = \{P_{\Theta} \equiv P(y|\Theta),  \Theta \equiv (\theta_{i})_{i=1}^{d} \in  {V}_{\Theta} \subset \Re^{d} \} \; ,
\end{eqnarray}
is given, i.e. the family of the probability distributions parameterized by $d$ non-random variables
$(\theta_{i})_{i=1}^{d}$ which are real-valued and belong to the parametric space  ${V}_{\Theta}$ of the parameter $\Theta$, i.e. $\Theta \in V_{\Theta} \subset \Re^{d}$.
Thus, the logarithm of the likelihood function  $\ln  P: V_{\Theta} \rightarrow \Re$ is defined on the space $V_{\Theta}$.

Let $\tilde{\Theta} \equiv (\tilde{\theta}_{i})_{i=1}^{d} \in V_{\Theta}$ be another value of the parameter or a value of the
estimator $\hat{\Theta}$ of the parameter $\Theta=(\theta_{i})_{i=1}^{d}$.
At every point, $P_{\Theta}$, the $d \times d$ - dimensional observed Fisher information (FI) matrix can be defined \cite{Pawitan,Dziekuje informacja_2}:
\begin{eqnarray}
\label{observed IF}
\texttt{i\!F}(\Theta) \equiv
- \, \partial^{i'} \partial^{i} \ln P(\Theta)
= \left( - \, \tilde{\partial}^{i'} \tilde{\partial}^{i} \ln P(\tilde{\Theta}) \right)_{|_{\widetilde{\Theta} = \Theta}} \,
\end{eqnarray}
and  $\partial^{i} \equiv \partial/\partial \theta_{i}$, $\tilde{\partial}^{i} \equiv \partial/\partial \tilde{\theta}_{i}$, $\,i,i'=1,2,..,d$.
%
%
It characterizes the local properties of
$P(y|\Theta)$. It is  symmetric and in field theory and statistical physics models with continuous, regular and normalized distributions, it is positively definite \cite{Pawitan}. We restrict the considerations to this case only.
The expected $d \times d$ - dimensional FI matrix on  ${\cal S}$ at point $P_{\Theta}$ is defined as follows \cite{Amari Nagaoka book}:
\begin{eqnarray}
\label{infoczekiwana}
I_F \left(\Theta\right) \equiv E_{\Theta} \left(\texttt{i\!F}(\Theta)\right) = \int_{\cal B} dy P(y|\Theta) \, \texttt{i\!F}(\Theta) \; ,
\end{eqnarray}
where the differential element
$dy \equiv d^{N}{\bf y} = d{\bf y}_{1} d{\bf y}_{2} ... d{\bf y}_{N}$.
The subscript $\Theta$ in the expected value
signifies the true value of the parameter under which the data $y$
are generated.
The FI matrix defines
on ${\cal S}$ the Riemannian Rao-Fisher metric \cite{Amari Nagaoka book,Dziekuje za channel}.
Sometimes, due to the probability distribution normalization and the regularity condition,  the $d \times d$ - dimensional observed Fisher information (FI) matrix can be
recorded in the "quadratic" form \cite{Amari Nagaoka book}:
\begin{eqnarray}
\label{observed IF Amari}
\texttt{i\!F} = \left( \partial^{i'} \! \ln P(\Theta)\;
\partial^{i}  \ln P(\Theta) \right) \; .
\end{eqnarray}
The central quantity of EPI analysis is the information channel capacity  $I$ which
is the trace of the (expected) Fisher information matrix. Since under above conditions,
the observed Fisher information matrix is diagonal $\texttt{i\!F}(\Theta)$ $= {\rm diag}(\texttt{i\!F}_{\!\!nn}(\Theta))$, hence
the information channel capacity $I(\Theta)$ is equal to:
\begin{eqnarray}
\label{def I dla diag IF}
I(\Theta) = \sum_{n=1}^{N}
\int_{\cal B} dy \, P(y|\Theta) \, \texttt{i\!F}_{\!\!nn}(\Theta)
=
\int_{\cal B} dy \, \textit{i} \; ,
\end{eqnarray}
where $\textit{i} :=  P(\Theta) \; \sum_{n=1}^{N} \texttt{i\!F}_{\!\!nn}(\Theta) $ is
the {\it information channel density} \cite{Dziekuje za channel,Dziekuje-za-EPR-Bohm}.

\section{Generalization of the Aoki-Yoshikawa  Model}

\label{epim}

The generalization of the AYM  presented in this paper consists in considering productivity as a  continuous random variable $A$. The transition from the discrete variable to the continuous one is performed via \begin{eqnarray}
\label{przyporzadkowanie Pa Pe}
A = a_{i} \rightarrow a \; \;\;\; {\rm and} \;\;\; p_{i}=\frac{n_{i}}{n}  \rightarrow p (a) \; .
\end{eqnarray}
As a consequence, the probability distribution function $p$ has to be normalized:
\begin{eqnarray}
\label{przyporzadkowanie Na Ne}
\sum_{i=1}^{g} p_{i} = 1 \rightarrow \int_{{\cal Y}_{a}} da \, p(a) =1  \; ,
\end{eqnarray}
where ${\cal Y}_{a}$ denotes the set of possible values of the productivity.
Analogously, the expectation value of the productivity is replaced in the following way
\begin{eqnarray}
\label{przyporzadkowanie SRa SRe}
\theta_{A} \equiv \left\langle A  \right\rangle = \sum_{i=1}^{g} p_{i} \, a_{i}   \rightarrow \int_{{\cal Y}_{a}} da \; p(a) \,a    \,  ,
\end{eqnarray}
with the constraint
\begin{eqnarray}
\label{wartosc oczekiwana produkcyjnosci}
\left\langle A  \right\rangle =  D/n \; .
\end{eqnarray}

In order to find the probability distribution of the level of productivity $A$ the EPI method is used. 
Below, the forms of the Fisher information will be adopted
to estimate the scalar parameter $\theta_{A}$, (\ref{przyporzadkowanie SRa SRe}),  \cite{Dziekuje-za-EPR-Bohm}:
\begin{eqnarray}
\label{srednia A}
\theta_{A} \equiv  \left\langle A  \right\rangle  = \int_{{\cal Y}_{a}} da \; p(a) \,a  \; ,
\end{eqnarray}
which in this paper is the expectation value of the random variable of the productivity level $A$.
The basic information on the Fisher information is given in Section~\ref{basic FI} and the construction of the information channel capacity can be found in \cite{frieden,Dziekuje za channel}.

Previous attempt to solve the Aoki-Yoshikawa productivity problem by the EPI method  was based on the Frieden-Soffer approach \cite{frieden}. However, this paper implements only the analytical form of the (modified) observed structural information principle \cite{Dziekuje-za-EPR-Bohm}.

\subsection{The expected Fisher information
and the information capacity of the channel~$(\theta_{A})$}

\label{expected Fisher information}

%

According to the {\it the main assumption of the EPI method}
proposed by Frieden and Soffer, the system itself
samples the space of positions (that is,
the space of values of productivity $A$ levels).
This space is accessible using its Fisherian degrees of freedom \cite{frieden,Dziekuje za channel}.
%
%
The sample of the EPI method (so-called inner sample \cite{Dziekuje za channel,Dziekuje-za-EPR-Bohm}) for the AY model is
$N=1$-dimensional\footnote{In
\cite{Frieden-Soffer,frieden}
the condition of the minimal value of
the information (kinematical) channel capacity
$I \rightarrow min$ is postulated as the one that fixes
the value of $N$ in a unique way.
However, sometimes the non-minimal values of $I$ are also discussed as they
lead to the EPI method's models, which are of a physical significance
\cite{Frieden-Soffer,frieden}. Some discussion on this topic can be also found in \cite{Dziekuje-za-EPR-Bohm}.
For example, from the analogy with the EPI model of momentum  distribution presented in \cite{frieden}, for $N>1$ the nonequilibrium, stationary solutions for the square flow (where the flow is proportional to the root productivity) might be obtained instead.
The point is that when choosing
the size $N$ of a sample, different classes of the EPI models are obtained. Yet, with the value of $N$ fixed,
the particular EPI model is provided by the solution of one or two information principles. In the case of the discussed AY model, the observed structural principle, which is a
differential one, and the variational principle are used.  The variational information principle is connected with the extremization of the physical information  (see Section~\ref{Informacja strukturalna EPR}).
Thus, the condition $I \rightarrow min$ is choosing the type of the model only \cite{frieden,Dziekuje informacja_1,Dziekuje-za-skrypt} whereas with the information principles, the particular solution inside this type of the model is found.
}.
Thus, both the sample space ${\cal B}$
and the base space of events are in AY case equivalent to ${\cal Y}_{a}$ \cite{Dziekuje-za-EPR-Bohm}.

As $N=1$ thus
$p(a|\theta_{A}) \equiv p(a)$
is the {\it likelihood function for the AY model} and the $d=1$-dimensional statistical space of for the AY model is as follows:
\begin{eqnarray}
\label{model statystyczny S dla AY}
{\cal S} = \{p_{\theta_{A}} \equiv p(a\,|\theta_{A}), \;  \theta_{A} \in  {V}_{\theta_{A}} \subset \Re \} \; ,
\end{eqnarray}
where ${V}_{\theta_{A}}$ is the parameter space of $\theta_{A}$.
%
%
%
For
$N=1$  the information channel
capacity $I$
reduces to the Fisher information
$I_{F}(\theta_{1}) = I_{F}(\theta_{A})$
for $\theta_{1}  \equiv \theta_{A}$, the only parameter.
The necessary steps leading from the general form of the Fisher information to the one used in this paper are similar as in \cite{Dziekuje-za-EPR-Bohm}.

{\it The probability amplitude} $q_{a} \equiv q\left(a|\theta_{A}\right)$ is defined in the following way \cite{Bengtsson_Zyczkowski,Amari Nagaoka book,Dziekuje-za-EPR-Bohm}:
\begin{eqnarray}
\label{inffishEPR2}
p\left(a\left|{\theta_{A}}\right.\right) = \frac{1}{4} \,
q^{2}\left(a|\theta_{A}\right) \;  .
\end{eqnarray}
The $N=1$-dimensionality of the sample means that the rank of the amplitude $q\left(a|\theta_{A}\right)$ of the (productivity) field is also equal to 1 \cite{frieden}.

As $\theta_{A}$ is the scalar parameter and the dimension of the
sample is equal to $N=1$
the information channel capacity $I({\theta_{A}})$
of the {\it measurement channel} $(\theta_{A})$ \cite{Dziekuje za channel} is equal to
{\it the Fisher information $I_{F}(\theta_{A})$ of the parameter} $\theta_{A}$ \cite{Dziekuje-za-EPR-Bohm}:
\begin{eqnarray}
\label{IF dla vartheta w AY}
I({\theta_{A}}) = I_{F}(\theta_{A})
\, .
\end{eqnarray}
$I_{F}(\theta_{A})$ is the information about the unknown parameter $\theta_{A}$ confined
in the $N=1$-dimensional sample for the random variable $A$.

The {\it analytical form}
of the (expected)
Fisher information on parameter $\theta_{A}$ is equal to \cite{Dziekuje-za-EPR-Bohm}:
\begin{eqnarray}
\label{postac I analytical AY}
I_{F}(\theta_{A}) &=&
\int_{{\cal Y}_{a}} {d a} \;
p\left(a\left|{\theta_{A}}
\right.\right) \texttt{i\!F}_{a}(\theta_{A})  \nonumber \\
&= &
\int_{{\cal Y}_{a}} {d a}\;
{p\left(a\left|{\theta_{A}}
\right.\right) \left( - \;\frac{\partial^{2}
\ln p \left(a\left|{\theta_{A}}\right.\right)}{\partial
\theta_{A}^{2}}\right)}  \nonumber
\\
&=&
\int_{{\cal Y}_{a}} {d a}\; \left( - \, \frac{\partial^{2} p\left(a\left|{\theta_{A}}\right.\right)}{\partial
\theta_{A}^{2}} \, +
{\frac{1}{p\left(a\left|{\theta_{A}}
\right.\right)}\left(\frac{\partial p\left(a\left|{\theta_{A}}\right.\right)}{\partial
\theta_{A}}\right)^{2}} \right) \nonumber \\
&=&
\int_{{\cal Y}_{a}} {d a}\;
\left( - \, \frac{\partial^{2}p\left(a\left|{\theta_{A}}\right.\right)}{\partial
\theta_{A}^{2}} \,
+
{\left( q_{a}^{'} \right)^{2}} \,
\right)  \nonumber \\
&=&
\int_{{\cal Y}_{a}} {d a}\;
\left( - \, q_{a} {
q_{a}^{''} } + \frac{\partial^{2}p\left(a\left|{\theta_{A}}\right.\right)}{\partial
\theta_{A}^{2}} \,
\right) \, ,
\end{eqnarray}
where Eq.(\ref{inffishEPR2}) and the denotations $q_{a}^{'} \equiv
\frac{dq_{a}(\theta_{A})}{d\theta_{A}}$ and $q_{a}^{''} \equiv
\frac{d^{2}q_{a}(\theta_{A})}{d\theta_{A}^{2}}$ have been used and the index in $\texttt{i\!F}_{a}(\theta_{A})$ signifies the dependence of the observed Fisher information on $a$.
In the last equality the relation:
\begin{eqnarray}
\label{P na q}
\frac{\partial^{2}p\left(a\left|{\theta_{A}}\right.\right)}{\partial
\theta_{A}^{2}} = \frac{1}{2} \,(q_{a}^{'})^{2} + \frac{1}{2} \, q_{a} \,q_{a}^{''}
\end{eqnarray}
was also applied.

Due to the normalization, (\ref{przyporzadkowanie Na Ne}):
\begin{eqnarray}
\label{normalization of p}
\int_{{\cal Y}_{a}} da \, p(a|\theta_{A}) =1  \; ,
\end{eqnarray}
and the regularity condition
\cite{Pawitan,Dziekuje-za-EPR-Bohm}:
\begin{eqnarray}
\label{regularity condition}
\!\!\!\!\! \int_{{\cal Y}_{a}} {d a}\;
\frac{\partial^{2}p\left(a|\theta_{A}\right)}{\partial \theta_{A}^{2}}  = \frac{\partial^{2} }{\partial \theta_{A}^{2}} \int_{{\cal Y}_{a}} {d a} \;  p\left(a|\theta_{A}\right)= \frac{\partial^{2} }{\partial \theta_{A}^{2}} 1 = 0 \; , \;
\end{eqnarray}
the {\it analytical form} (\ref{postac I analytical AY}) of the Fisher information transforms into the following {\it metric form} \cite{Dziekuje-za-EPR-Bohm}:
\begin{eqnarray}
\label{inffishEPR-origin}
\!\!\!\!\!\!\!\! \!\!\!\!\!\!\!\!
I_{F}(\theta_{A})  &=&  \int_{{\cal Y}_{a}} {d a} \;
p\left(a\left|{\theta_{A}}
\right.\right) \widetilde{\texttt{i\!F}}_{a}(\theta_{A})
\nonumber \\
\!\!\!\!\!\!\!\! &=&
\int_{{\cal Y}_{a}} {d a} \;
{\frac{1}{p\left(a\left|{\theta_{A}}
\right.\right)}\left(\frac{\partial p\left(a\left|{\theta_{A}}\right.\right)}{\partial
\theta_{A}}\right)^{2}}  =
\int_{{\cal Y}_{a}} {d a} \; {\left(q_{a}^{'} \right)^{2}} \; .
\end{eqnarray}

Due to Eq.(\ref{regularity condition}) and in accordance with Eq.(\ref{IF dla vartheta w AY}),
 both the {\it EPI method form} of the (expected) Fisher information \cite{Dziekuje-za-EPR-Bohm} for the AYM and its information channel capacity for the measurement channel $(\theta_{A})$ are equal to:
\begin{eqnarray}
\label{inffishEPR}
\;\;\;\;\;\;\;\;
I({\theta_{A}}) = I_{F}(\theta_{A})
= -  \int_{{\cal Y}_{a}} {d a} \; q_{a}(\theta_{A}) q_{a}^{''}(\theta_{A})    \; .
\end{eqnarray}

{\it The information channel capacity $I$ is the one which
enters into the estimation procedure of
the} EPI {\it method}.
The presented below
derivation of the form of amplitude $q_{a}$ as the self-consistent solution of the
information principles
consistently uses the analytical form \cite{Dziekuje informacja_2,Dziekuje-za-EPR-Bohm} of the
structural information principle, and in this respect, it is different from the derivation of the Boltzmann distribution given in~\cite{frieden}.

\subsection{The information principles and generating equation}

\label{Informacja strukturalna EPR}


In \cite{Dziekuje informacja_1} the existence of the (total)
{\it physical information}\footnote{With such general understanding of $K$,
the diversity of the equations of the EPI method is a consequence
of diverse preconditions dictated by the investigated phenomenon
\cite{Frieden-Soffer,frieden,PPSV,Dziekuje informacja_2}).
}
$K$
\begin{eqnarray}
\label{physical K}
K = I + Q  \geq 0 \;
\end{eqnarray}
was postulated\footnote{See also \cite{Dziekuje informacja_2,Dziekuje za channel,Dziekuje-za-EPR-Bohm}, where the differences between the Frieden-Soffer original form of the physical information and information principles used in this paper are discussed.
}.
The choice of the intuitive condition $K \geq 0$ is connected with the
{\it expected structural information principle} of the EPI method:
\begin{eqnarray}
\label{condition from K}
I + \kappa \, Q  = 0 \;
\end{eqnarray}
derived for $\kappa=1$ in \cite{Dziekuje informacja_2},
where $\kappa$ is the so-called efficiency coefficient introduced  in \cite{frieden}.
The general forms of
$I$ and
the {\it structural information} $Q$ are given in \cite{frieden,Dziekuje informacja_2}.
The form of the information principle, which is more fundamental than
(\ref{condition from K}), is the {\it observed structural information
principle} that has the form $ \texttt{q\!F} + \texttt{i\!F} = 0$ \cite{Dziekuje informacja_2}.
The derivation of the particular form of the observed structural information
principle for the AY model is similar to the one for the EPR-Bohm problem \cite{Dziekuje-za-EPR-Bohm} so only necessary steps will be presented.

The other information principle of the EPI method is the {\it variational information principle} \cite{frieden}:
\begin{eqnarray}
\label{variational inf princ general}
\delta(I + Q) = 0 \;  .
\end{eqnarray}
It has to be stressed that it is the (modified) observed structural information principle \cite{Frieden-Soffer,frieden}, \cite{Dziekuje-za-EPR-Bohm,Dziekuje-za-skrypt}, and not the expected one, which is solved self-consistently together with the variational information principle. Below, both information principles will be constructed and solved in case of the AYM.

{\bf Note.} To obtain the value of the efficiency coefficient $\kappa$, the information principles, i.e., the (modified) observed structural information
principle \cite{Dziekuje-za-EPR-Bohm,Dziekuje-za-skrypt} and the variational information principle, together with the physical preconditions which are specific for the model, e.g., some symmetry conditions, have to be solved simultaneously \cite{frieden}. As a result both, the specific form of $Q$ and the value of $\kappa$ are obtained \cite{frieden}. In \cite{frieden} it is suggested that for the EPI models which have the quantum counterparts, $\kappa=1$ \cite{Dziekuje-za-EPR-Bohm}, whereas for the classical models $0 \leq \kappa \leq 1$ \cite{frieden}.  From the below analysis it follows that in the Aoki-Yoshikawa model $\kappa=1$, analogously as in the EPI model of the
Boltzmann energy distribution \cite{frieden}.  From the analysis presented in \cite{Dziekuje-za-EPR-Bohm} it also follows that Frieden's EPI method of ``coverage of quantum mechanics'' can be constructed by giving the quantum mechanical interpretation to the statistical probability amplitudes.
Yet,
in the case of the Einstein-Podolsky-Rosen-Bohm problem the quantum character of the amplitudes is provided by the EPI statistical information theory modelling itself \cite{Dziekuje-za-EPR-Bohm}.
Thus, the amplitudes $q_{a}(\theta_{A})$ of the EPI method for the Aoki-Yoshikawa model could gain the quantum mechanical interpretation if only a reason for the quantization (e.g., as $a_i = i a_0$ in Eq.(\ref{ciag a i})) of the productivity levels exists,
e.g., as in the Einstein model for specific heat \cite{Einstein_model}.
The assumption (\ref{ciag a i}) is used in \cite{garibaldi} for the AYM, which is quoted only in order to compare this model with the one obtained via the EPI method in Section~\ref{porownanie}.
%

\subsubsection{The information principles for the Aoki-Yoshikawa model}

\label{EPR-Bohm information principles}

%
%
The structural information
$Q$ \cite{Dziekuje informacja_2}
in the AYM
for the system described by the set of amplitudes $q_{a}$ \cite{Dziekuje-za-EPR-Bohm} is as follows:
\begin{eqnarray}
\label{strukturalnaEPR}
Q  \equiv \frac{1}{4}
\int_{{\cal Y}_{a}} {d a} \; {q_{a}^{2}(\theta_{A}) \, \texttt{q\!F}_{a}(q_{a})} \; ,
\end{eqnarray}
where for the simplicity reason the denotation $q_{a}\equiv q_{a}(\theta_{A}) \equiv q(a|\theta_{A})$ is used.

Now, the physical information $K$, (\ref{physical K}),
in the AYM is as follows
\cite{frieden,Dziekuje informacja_2}:
\begin{eqnarray}
\label{TPI diag}
K = I + Q = \int_{{\cal Y}_{a}} {d a} \; k_{a}(\theta_{A})  \; ,
\end{eqnarray}
where $I$ is
given by Eq.(\ref{postac I analytical AY}).
In Eq.(\ref{TPI diag}), $k_{a}(\theta_{A})$ is the {\it density of the
physical information}, which according to Eqs. (\ref{postac I analytical AY}), (\ref{strukturalnaEPR}) and
(\ref{P na q}) is equal to
\begin{eqnarray}
\label{k EPR}
k_{a}(\theta_{A}) &=&   -  \, q_{a} {
q_{a}^{''} } + \frac{\partial^{2} p\left(a\left|{\theta_{A}}\right.\right)}{\partial
\theta_{A}^{2}}  + \frac{1}{4} \, q_{a}^{2} \,  \texttt{q\!F}_{a}(q_{a})    \nonumber \\
&=&   - \, \frac{1}{2} \, q_{a} { q_{a}^{''} } + \frac{1}{2} (q_{a}^{'})^2
+ \frac{1}{4} \, q_{a}^{2} \,  \texttt{q\!F}_{a}(q_{a})    \nonumber \\
&=&   - \frac{1}{2} \, q_{a} {
q_{a}^{''} }  + \frac{1}{4} \, q_{a}^{2} \, \widetilde{\texttt{q\!F}}_{a}(q_{a})   \; ,
\end{eqnarray}
where the {\it modified observed structural information} $\widetilde{\texttt{q\!F}}_{a}$ used in the EPI method has been introduced \cite{Dziekuje-za-EPR-Bohm}:
\begin{eqnarray}
\label{qF in EPI}
\;\;\; \widetilde{\texttt{q\!F}}_{a}(q_{a})  := \frac{2}{q_{a}^{2}(\theta_{A})}
(q_{a}^{'})^{2}  +  \texttt{q\!F}_{a}(q_{a}) \; .
\end{eqnarray}

Under the assumption of analyticity of the log-likelihood function $\ln p\left(a\left|{\theta_{A}}\right.\right)$,
the Taylor expansion $\ln p\left(a\left|{\tilde{\theta}_{A}}\right.\right)$
around the true value of $\theta_{A}$ with the use of the denotations
 $\frac{\partial^{2} \ln p(\theta_{A})}{\partial \theta_{A}^{2}} \equiv \frac{\partial^{2} \ln p(\tilde{\theta}_{A})}{\partial \tilde{\theta}_{A}^{2}}\mid_{\tilde{\theta}_{A}=\theta_{A}}$ and
$q_{a}^{'}(\theta_{A}) \equiv \frac{\partial q_{a}(\tilde{\theta}_{A})}{\partial \tilde{\theta}_{A}}\mid_{\tilde{\theta}_{A}=\theta_{A}}$,
leads to the following form of the observed structural information \cite{Dziekuje-za-EPR-Bohm}:
\begin{eqnarray}
\label{Freiden like equation with qF modyfikowane}
\texttt{q\!F}_{a}(q_{a})   =  \frac{1}{q_{a}^{2}(\theta_{A})}  2 \, \left( q_{a}(\theta_{A}) q_{a}^{''}(\theta_{A})   -   \, (q_{a}^{'}(\theta_{A}))^{2} \right) \; .
\end{eqnarray}

Here the appearance of $q_{a}$ in the argument of $\texttt{q\!F}_{a}$ means that the probability  $p(a|{\tilde{\theta}_{A}})$ (and its derivatives) present  in $\texttt{q\!F}_{a}$ in the derivation \cite{Dziekuje-za-EPR-Bohm} of (\ref{Freiden like equation with qF modyfikowane}),
has been replaced by the amplitude $q_{a}$ (and its derivatives).

In what follows, the forms of the amplitudes $q_{a}$ that are the solution to the AYM will be searched for among combinations of the exponential functions.
%
Additional assumption that the term with the first derivative $q_{a}^{'}(\theta_{A})$ on the RHS of the above equation cancels with a term in $\texttt{q\!F}_{a}(q_{a})$ has been made.

Now, after moving the term $\frac{1}{2}\, (q_{a}^{'})^{2}$ in Eq.(\ref{Freiden like equation with qF modyfikowane}) from the Fisher information part on its RHS to the structural one on its LHS the {\it modified observed structural information principle} was  obtained \cite{Dziekuje-za-EPR-Bohm}. (This shift between $\texttt{q\!F}_{a}(q_{a})$ and $\widetilde{\texttt{q\!F}}_{a}(q_{a})$  is then used in  Eq.(\ref{k EPR}).) Thus, the modified observed structural information principle for the AYM has the following form \cite{Dziekuje-za-EPR-Bohm}:
\begin{eqnarray}
\label{mikroEPR}
&-&  2 \, q_{a}(\theta_{A}) q_{a}^{''}(\theta_{A}) + q_{a}^{2}(\theta_{A}) \, \widetilde{\texttt{q\!F}}_{a}(q_{a})  = 0  \;   ,
\end{eqnarray}
where
\begin{eqnarray}
\label{qF modyfikowane}
\;\;\;\; \widetilde{\texttt{q\!F}}_{a}(q_{a}) &\equiv&
\left( \texttt{q\!F}_{a}(q_{a}) +  \frac{1}{q_{a}^{2}(\theta_{A})} \, 2  \, (q_{a}^{'}(\theta_{A}))^{2} \right)
\nonumber \\
&=& \frac{1}{q_{a}^{2}(\theta_{A})}  2 \, q_{a}(\theta_{A}) q_{a}^{''}(\theta_{A})   \; .
\end{eqnarray}
Equation (\ref{mikroEPR}) arises purely as a result of analyticity of the log-likelihood function.

The LHS of Eq.(\ref{mikroEPR}) is (up to the factor $\frac{1}{4}$) the density of the
physical information $k_{a}(\theta_{A})$ given by Eq.(\ref{k EPR}).
This one is the function of the observed structural information $\texttt{q\!F}_{a}(q_{a})$
(which  at most can be the function of the amplitudes $q_{a}(\theta_{A})$), of the amplitudes themselves $q_{a}(\theta_{A})$
and of their second derivatives.

{\bf Comment}:
In the AYM, the efficiency factor $\kappa$ is equal
to $\kappa =1$ \cite{frieden}. This follows from the fact that except for the
information principles, no additional differential constraints are  put upon the amplitudes $q_{a}$.
Thus, the presented EPI model is a pure analytic one \cite{Dziekuje informacja_2}, similarly in this respect as in the EPR-Bohm problem~\cite{Dziekuje-za-EPR-Bohm}.
\\

Using Eq.(\ref{k EPR}) the physical information $K$, (\ref{TPI diag}),  takes the following form:
\begin{eqnarray}
\label{K analytical in EPR}
K &=& I + Q
\\
&=& \int_{{\cal Y}_{a}} {d a} \;  \left(- \, \frac{1}{2} \, q_{a} {
q_{a}^{''} }  + \frac{1}{4} \, q_{a}^{2}(\theta_{A}) \, \widetilde{\texttt{q\!F}}_{a}(q_{a}) \right) \, \; . \nonumber
\end{eqnarray}
From Eq.(\ref{mikroEPR}) the expected structural information principle (see Eq.(\ref{condition from K})), for $\kappa=1$, follows:
\begin{eqnarray}
\label{expect structural IP in EPR}
I+Q=0 \; ,
\end{eqnarray}
where $I+Q$ is given by the RHS of Eq.(\ref{K analytical in EPR}).

The differential equation (\ref{mikroEPR}) is the first one from the
information principles used in the EPI method.
The second one presented below is the variational information principle
\cite{frieden,Dziekuje informacja_1,Dziekuje informacja_2,Dziekuje-za-EPR-Bohm}.

%
%
In order to obtain the variational information principle, we have to transform the physical information $K$, (\ref{K analytical in EPR}), into the {\it metric form}, i.e., the one quadratic in $q_{a}^{'}$.
Therefore, after integration by parts, $K$ can be rewritten as follows \cite{Dziekuje-za-EPR-Bohm}:
\begin{eqnarray}
\label{K variational in EPR}
K = I + Q = \int_{{\cal Y}_{a}} {d a} \, \left( k_{a}^{met}(\theta_{A}) - \frac{\textsc{c}_{a}}{2} \,   \right)  \; ,
\end{eqnarray}
where the constant $\textsc{c}_{a}$ is equal to:
\begin{eqnarray}
\label{stalaCEPR male} \textsc{c}_{a} =
\left({q_{a}\left({\infty}\right)q_{a}^{'}
\left({\infty}\right) - q_{a}\left(a_{0}\right)q_{a}^{'}\left(a_{0}\right)}\right)
\; ,
\end{eqnarray}
where $a_{0}$ is the smallest (absolute) level
of the productivity
and $k_{a}^{met}(\theta_{A})$ is the {\it metric form} of
density of the physical information:
\begin{eqnarray}
\label{k met variational in EPR}
k_{a}^{met}(\theta_{A}) = \frac{1}{2} \, q_{a}^{'2}  + \frac{1}{4} \,  q_{a}^{2}(\theta_{A}) \, \widetilde{\texttt{q\!F}}_{a}(q_{a}) \; .
\end{eqnarray}

{\it The variational information principle} has the form \cite{frieden,Dziekuje-za-EPR-Bohm}:
\begin{eqnarray}
\label{zskalarnaEPR}
\delta_{(q_{a})} K &\equiv&
\delta_{(q_{a})}\left( I + Q\right) = \\
&=&
\delta_{(q_{a})} \! \left(\, \int_{{\cal Y}_{a}} {d a} \,
{ ( \, k_{a}^{met}(\theta_{A}) - \frac{\textsc{c}_{a}}{2} \, ) \, } \right)  = 0 \; . \nonumber
\end{eqnarray}
The solution of the {\it variational} problem (\ref{zskalarnaEPR})
with respect to $q_{a}$ is the {\it Euler-Lagrange equation}:
\begin{eqnarray}
\label{row E-L dla EPR}
\frac{d}{d\theta_{A}}\left(\frac{\partial
k_{a}^{met}(\theta_{A})}{\partial
q_{a}^{'}(\theta_{A})}\right)
&=&
\frac{\partial
k_{a}^{met}(\theta_{A})}{\partial q_{a}} \;   .
\end{eqnarray}
From  this equation and for $k_{a}^{met}(\theta_{A})$ as in Eq.(\ref{k met variational in EPR}),
the following differential equation is obtained for every amplitude $q_{a}$:
\begin{eqnarray}
\label{rozweularaEPR}
q_{a}^{''}=\frac{1}{2}\frac{{d (\frac{1}{2} q_{a}^{2}  \widetilde{\texttt{q\!F}}_{a}(q_{a}))}}{{dq_{a}}} \;  .
\end{eqnarray}
As $q_{a}^{2}(\theta_{A}) \widetilde{\texttt{q\!F}}_{a}(q_{a})$ is
explicitly the function of $q_{a}$ only, the total
derivative has replaced the partial derivative over $q_{a}$ present in
Eq.(\ref{row E-L dla EPR}). The obtained form of equation (\ref{rozweularaEPR}) differs slightly from the Frieden form \cite{frieden} and is the same as in \cite{Dziekuje-za-EPR-Bohm}.
The origin of this difference is the fully analytical form of density of the physical information (\ref{k EPR}).

The modified observed structural information principle
(\ref{mikroEPR}) and the variational information principle
(\ref{zskalarnaEPR}) (from which the Euler-Lagrange
equation (\ref{rozweularaEPR}) follows)
serve for the derivation of the equation which generates the distribution.

\subsubsection{The derivation of the generating equation}

\label{derivation of the generating equation}

\vspace{3mm}

Using the relation (\ref{rozweularaEPR}) in Eq.(\ref{mikroEPR}), one can obtain:
\begin{eqnarray}
\label{generating eq general}
\frac{1}{2}q_{a}\frac{{d( q_{a}^{2}
\widetilde{\texttt{q\!F}}_{a}(q_{a}))}}{{dq_{a}}}
=  q_{a}^{2}
\widetilde{\texttt{q\!F}}_{a}(q_{a}) \;  .
\end{eqnarray}
The above equation can be rewritten in a handier form:
\begin{eqnarray}
\frac{{2dq_{a}}}{{q_{a}}}=\frac{{d \left(\frac{1}{2} q_{a}^{2}
\widetilde{\texttt{q\!F}}_{a}(q_{a})\right)}}{{\frac{1}{2} q_{a}^{2}
\widetilde{\texttt{q\!F}}_{a}(q_{a})
}}
\; ,
\end{eqnarray}
from which, after integration on both sides, the following results can be obtained:
\begin{eqnarray}
\label{jEPR}
& & \!\!\!\!\!\!\!
\frac{1}{2} q_{a}^{2}(\theta_{A})
\widetilde{\texttt{q\!F}}_{a}(q_{a}) = \alpha^2 \; q_{a}^{2}(\theta_{A})
\nonumber \\
& & \!\!\!\!\!\!\!
{\rm hence} \;\;\;
\widetilde{\texttt{q\!F}}_{a}(q_{a}) = 2 \, \alpha^2  \; ,
\end{eqnarray}
where the constant of integration $\alpha^{2}$ is a complex number in general.
%
By substituting Eq.(\ref{jEPR}) into
Eq.(\ref{rozweularaEPR}), we obtain the searched for differential
{\it generating equation} for the amplitudes  $q_{a}$ \cite{frieden}:
\begin{eqnarray}
\label{row generujace dla amplitud w EPR}
q_{a}^{''}(\theta_{A}) &=& \alpha^2 \;q_{a}(\theta_{A}) \;\;\; {\rm for} \;\;\;
\theta_{A} \in V_{\theta_{A}} \; ,
\end{eqnarray}
which is the consequence of both information principles - the structural and variational ones.
This result was obtained previously in
\cite{frieden} for the
Boltzmann probability distribution
but the arrival at the structural information principle
is here \cite{Dziekuje-za-EPR-Bohm} different and the form of both information principles also differs slightly.

{\bf Note.}
If
an explicit dependence of $\widetilde{\texttt{q\!F}}_{a}(q_{a})$ on the productivity $a$
is assumed, i.e. $\widetilde{\texttt{q\!F}}_{a}(q_{a}, a)$,  then a wider scope of solutions to the problem (\ref{generating eq general}) is possible, which also includes non-equilibrium solutions  \cite{frieden}. These solutions correspond to the
non-equal probability
of the occupation of the particular $s$th sector by the $i$th
worker for all particular configurations of these occupations (contrary to the assumption used in (\ref{prawdopodobienstwo pojawieniem sie wektora n})).

\section{The probability distribution for the Aoki-Yoshikawa model}

\label{probability distribution for Aoki-Yoshikawa model}

\subsection{The definition of the variable of the additive fluctuations}

\label{variable of the additive fluctuations}

The EPI method analysis for the distribution of the level of productivity is in accord with the general approach of Frieden.
The displacement $X_{a}$, defined as $X_{a} = A - \left\langle A  \right\rangle$, is used instead of
values of the productivity level $A$.
Thus, the additive partition is performed: $Y_{a} \equiv A = \left\langle A  \right\rangle + X_{a}$ (similarly, as for the Boltzmann distribution in \cite{frieden}).
It can be performed at the level of the information channel capacity, as it was originally proposed in   \cite{frieden} and developed in \cite{Dziekuje za channel} for the general distribution which is free of necessity to set the requirement for the shift-invariance, or it can be made at the level of the generating equation.  The latter possibility has been chosen as in the considered case this is the simple one, i.e.:
\begin{eqnarray}
\label{przyporzadkowanie xa}
{\bf y}_{a} \equiv a =  \theta_{A}   + {\bf x}_{a} \; , \;\;\;\; a_{0} \le {\bf y}_{a} \le \infty \; , \;\;\;  {\bf x}_{a}^{min} = a_{0} - \theta_{A} \le {\bf x}_{a} < \infty\; ,
\end{eqnarray}
where $X_{a} = {\bf x}_{a}$ is a particular displacement.
The simplifying assumption that the fluctuation of productivity is unbounded from above, is used (compare \cite{frieden} for the discussion on distribution of the energy fluctuation). Then, the EPI model is built over the space ${\cal X}_{a}$ of the displacements ${\bf x}_{a}$, which in our case is $\Re$ \cite{Dziekuje za channel}.

A simplifying notation now will be introduced:
\begin{eqnarray}
\label{zapis dla qn w xn}
q_{\theta_{A}}({\bf x}_{a}) \equiv q({\bf x}_{a} + \theta_{A}|\theta_{A}) = q(a|\theta_{A})  \;,
\end{eqnarray}
which {\it leaves the whole information on  $\theta_{A}$ that characterizes $q({\bf x}_{a} + \theta_{A}|\theta_{A})$ in the index of the amplitude $q_{\theta_{A}}({\bf x}_{a})$} (and similarly for the original distribution  $p_{\theta_{A}}({\bf x}_{a}) \equiv  p({\bf x}_{a}+\theta_{A}|\theta_{A}) = p(a|\theta_{A})$). \\

Now, appealing to the ``chain rule'' for the derivative:
\begin{eqnarray}
\label{chain rule}
\frac{d}{d {\bf \theta_{A}}} =  \frac{d (a -
\theta_{A})}{d \theta_{A}} \, \frac{d}{d (a -
\theta_{A})}  = - \; \frac{d}{d (a - \theta_{A})} = - \; \frac{d}{d\,{\bf x}_{a}} \;
\end{eqnarray}
a transfer from the statistical form (\ref{row generujace dla amplitud w EPR}) of the generating equation to its {\it kinematical form}\footnote{
Note: Taking into account that $d{\bf x}_{a} = d{\bf y}_{a}$, which is connected with the fact that parameter $\theta_{A}$ is a constant, we can transfer from the statistical form of the physical information $K = I + Q$ (\ref{TPI diag})-(\ref{k EPR}) to its {\it kinematical form} with the information channel capacity as follows \cite{frieden,Dziekuje informacja_2,Dziekuje-za-EPR-Bohm}:
\begin{eqnarray}
\label{I analytical in EPR kinematical}
I = \int_{{\cal X}_{a}} {d {\bf x}_{a}} \;   \left(
\frac{d q_{\theta_{A}}({\bf x}_{a})}{d\,{\bf x}_{a}} \right)^{2}   \;
\end{eqnarray}
and the structural information in the form:
\begin{eqnarray}
\label{Q analytical in EPR kinematical}
Q =  \int_{{\cal X}_{a}} {d {\bf x}_{a}} \;  \left( \frac{1}{4} \, q_{\theta_{A}}({\bf x}_{a})^{2} \, \texttt{q\!F}_{a}(q_{\theta_{A}}({\bf x}_{a})) \right) \; .
\end{eqnarray}
}:
\begin{eqnarray}
\label{produkcyjnosc row generujace}
\frac{d^{2} q_{\theta_{A}}({\bf x}_{a})}{d\,{\bf x}_{a}^2} = \alpha^{2} \, q_{\theta_{A}}({\bf x}_{a}) \; ,
\end{eqnarray}
where $q_{\theta_{A}}({\bf x}_{a})$ is the amplitude of the distribution of the productivity level fluctuation and it was chosen $\alpha$ to be a real constant (see footnote~\ref{alfa imaginary}).

\subsection{The solution of the generating equation}

\label{The solution of the generating equation}

As the amplitude $q_{\theta_{A}}$ is a real one, thus $\alpha^{2}$ in Eq.(\ref{produkcyjnosc row generujace}) has also to
be real. When the value of the fluctuation of the productivity ${\bf x}_{a}$ is not bounded from above,
and this condition is realized by ${\bf x}_{a}^{max}$ approaching infinity, then $\alpha$ has to be  real. In this case from the normalization of the squared amplitude we get
\begin{eqnarray}
\label{unormowanie q2}
\frac{1}{4} \int_{{\bf x}_{a}^{min}}^{\infty} d {\bf x}_{a} \, q_{\theta_{A}}^{2}({\bf x}_{a})  = \int_{{\bf x}_{a}^{min}}^{\infty}{ d{\bf x}_{a} \, p_{\theta_{A}}({\bf x}_{a})} = 1 \; ,
\end{eqnarray}
it follows that the solution of Eq.(\ref{produkcyjnosc row generujace}) is purely of an {\it exponential}
character\footnote{The
other possibility for $\alpha$ is that it is a purely imaginary number. Then the solution has the
trigonometric character \cite{frieden}. Yet, when ${\bf x}_{a}^{max} \rightarrow \infty$ then due to the normalization condition (\ref{unormowanie q2}), the trigonometric solution is not the allowed one.
(The case when the parametric space is finite can lead to the trigonometric solution, as it is in case of the EPR-Bohm problem \cite{frieden,Dziekuje-za-EPR-Bohm}).
\label{alfa imaginary}}
\cite{frieden}:
\begin{eqnarray}
\label{exponential}
q_{\theta_{A}}({\bf x}_{a}) =
B \, \exp\left(- \alpha \, {\bf x}_{a}\right) +
C \, \exp\left(\alpha \, {\bf x}_{a}\right) \; , \quad \alpha \in \mathbf{R}_{+} \; ,
\end{eqnarray}
where $B$ and $C$ are real constants.
%
As the normalization condition (\ref{unormowanie q2}) is defined in the interval ${\bf x}_{a}^{min} \le {\bf x}_{a} < \infty$ thus, the part of the solution with the positive exponent has to be rejected due to its divergence to infinity. Therefore, the requirement that $\alpha > 0$ leads to $C=0$.

In summary, the searched form of the amplitude is as follows:
\begin{eqnarray}
\label{qn rozwiazanie dla E}
q_{\theta_{A}}({\bf x}_{a}) = {B \exp\left({-\alpha \, {\bf x}_{a}}\right)}\; , \quad  \alpha\in\mathbf{R}_{+} \; , \quad {\bf x}_{a}^{min} \le {\bf x}_{a} < \infty \; .
\end{eqnarray}
From this and from the normalization condition  (\ref{unormowanie q2}), the constant $B$ is obtained:
\begin{eqnarray}
B = \pm \,2 \,\sqrt{2 \alpha }\,\exp\left({\alpha \, {\bf x}_{a}^{min}}\right)\; .
\end{eqnarray}
Thus, the final form of the amplitude  for $\alpha \in \mathbf{R}_{+}$ has the form
\begin{eqnarray}
\label{rozw cosinus dla ampl dla A}
q_{\theta_{A}}({\bf x}_{a}) = \pm \,2 \,\sqrt{2 \alpha}\,\exp\left[{\alpha \left({{\bf x}_{a}^{min}-{\bf x}_{a}}\right)}\right] \; ,\quad \alpha \in \mathbf{R}_{+} \;
\end{eqnarray}
and $\alpha$ is given in units $\left[1/productivity\right]$.

The following probability distribution of the fluctuation of productivity level  ${\bf x}_{a}$ can be  obtained from the amplitude (\ref{rozw cosinus dla ampl dla A}):
\begin{eqnarray}
\label{pn dla N=1}
p\left({\bf x}_{a}\right) = \frac{1}{4} \, q^{2}\left({\bf x}_{a}\right) = 2  \alpha\, \exp\left[{2\alpha\left({{\bf x}_{a}^{min} - {\bf x}_{a}}\right)}\right] \; , \quad \alpha \in \mathbf{R}_{+} \; .
\end{eqnarray}

In accordance with Eq.(\ref{przyporzadkowanie xa}) the following relation holds ${a} = \theta_{A} + {\bf x}_{a}$ thus, $d{a}/d{\bf x}_{a} = 1$. Therefore, the distribution of the random variable $A$ has the form:
\begin{eqnarray}
\label{rozklad p od E}
p\left(a\right)  = p\left({\bf x}_{a}\right) \frac{1}{|d a/d{\bf x}_{a}|} = 2 \alpha \,\exp \left[{-2\alpha\left({a - a_{0}}\right)}\right] \; , \quad a_{0} \le a <\infty \; .
\end{eqnarray}
Now, as the expectation value of $A$ is equal to\footnote{
Let us notice that from Eq.(\ref{srednia E}) it follows that $A$ is the unbiased estimator of the expectation value $\langle {A} \rangle$ of the level of productivity, i.e., $\widehat{\langle A \rangle} = A$.
}:
\begin{eqnarray}
\label{srednia E}
\langle {A} \rangle \equiv \theta_{A} = \int\limits_{a_{0}}^{+\infty}{d a \, p\left(a\right) a} \; ,
\end{eqnarray}
thus, inserting Eq.(\ref{rozklad p od E}) into (\ref{srednia E}), we obtain:
\begin{eqnarray}
\label{stala alfa}
2\, \alpha =  \left(\langle A \rangle - a_{0}\right)^{-1} \; .
\end{eqnarray}
\begin{figure}[t]
\begin{center}
\includegraphics[width=142mm,height=80mm]{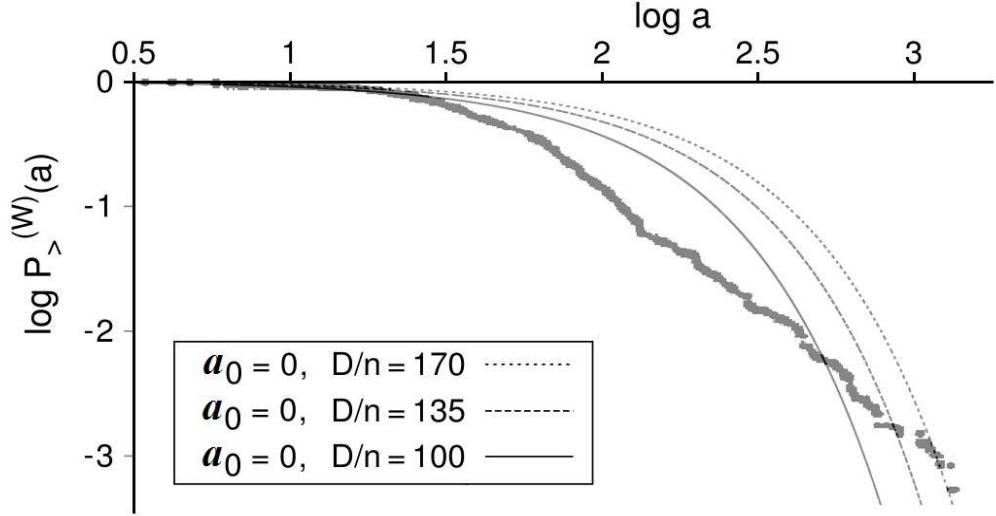}
\end{center}
\hspace{10mm}
\caption{This figure depicts the comparison of the cumulative probability distributions $P^{({\rm W})}_{>}({\rm a})$
$= \int_{{\rm a}}^{\infty} p\left( a \right) d a$
\cite{Labour-Prod} calculated for the
probability distribution of the level of productivity given by Eq.(\ref{rozklad koncowy A}) with the one observed  across workers (W) [source
\cite{Labour-Prod} for the 2005 year's data] (thick dotted line). The productivity cut ``${\rm a}$'' is given in the unit of $10^6$ yen/person \cite{Labour-Prod} and
on both axis the common logarithmic scale is used.
The smallest absolute level of the productivity $a_{0}$ of the worker is taken to be equal to zero. The ratio $D/n$ (which value is equal to the Boltzmann temperature \cite{Labour-Prod}) has its central value equal to 135 (long dashed line). For comparison, two additional curves, the left one with $D/n = 100$ (solid line) and the right one with $D/n = 170$ (short dashed line) are also plotted \cite{Labour-Prod}.
}
\vspace{-2mm}
\label{fig-1}
\end{figure}

Taking into account the constraint $\left\langle A  \right\rangle =  D/n$, (\ref{wartosc oczekiwana produkcyjnosci}), using Eq.(\ref{rozw cosinus dla ampl dla A}) finally  the amplitude can be obtained:
\begin{eqnarray}
\label{rozw dla ampl dla qA}
q_{\theta_{A}}({\bf x}_{a}) &=& \pm \, 2 \, \frac{1}{\sqrt{ D/n  - a_{0} \,}}\, \exp\left[ -\,  {\frac{{\bf x}_{a} + \left( D/n - a_{0}\right)}{2 \left( D/n  - a_{0}\right)} }\right] \; , \\
& & {\rm where} \;\;\;\; a_{0} - D/n \le {\bf x}_{a} < \infty\; \nonumber
\end{eqnarray}
and the search for probability distribution of the level of productivity:
\begin{eqnarray}
\label{rozklad koncowy A}
p\left( a \right) = \left\{ \begin{array}{l}
\frac{1}{(D/n) - a_{0}} \; \exp\left( - \, \frac{a - a_{0}}{(D/n) - a_{0}} \right) \;\;\;  \;\;\;\;\;\;  {\rm for} \quad \;\;\; a \ge a_{0} \\
\quad \quad 0  \;\,    \quad \quad \quad \quad \quad  \quad \quad \quad \quad \quad  \;\;\; {\rm for} \quad  \;\;\; a < a_{0} \end{array} \right. \; .
\end{eqnarray}
The distribution (\ref{rozklad koncowy A}) is the final result of the EPI method for AY model of productivity.

Finally, Figure~1 shows the comparison of the cumulative probability distributions $P^{({\rm W})}_{>}({\rm a}) = \int_{{\rm a}}^{\infty} p\left( a \right) d a$ considered in \cite{Labour-Prod} for the
probability distribution of the level of productivity given by Eq.(\ref{rozklad koncowy A}) with the observed productivity distribution across workers (W)
(source
\cite{Labour-Prod}).
If one considers the whole range of the productivity cut ``${\rm a}$''
then the exponential law (\ref{rozklad koncowy A}) (with the smallest absolute level of the productivity $a_{0}$ of the worker equal to zero) fits the data reasonably well.

\section{Comparison of two approaches}


\label{porownanie}

To compare the approaches, the discrete variable and assumptions have to be reproduced. The assumptions that have been made in the AYM are the following (\ref{ciag a i}):
\begin{eqnarray}
a_{i} = i \; a_{0} \; \;\;\; {\rm where} \;\;\; i=1,2,...,g \;,
\end{eqnarray}
the number of production sectors is great, $g >> 1$, and
\begin{eqnarray}
\label{defin r}
r \equiv \frac{D/n}{a_{0}} \; .
\end{eqnarray}
This leads to:
\begin{eqnarray}
P(i|{\bf n}^{*}) = \frac{n_{i}^{*}}{n} \approx \frac{1}{r-1} \left( \frac{r-1}{r} \right)^{i}\approx (\frac{1}{r} + \frac{1}{r^2}) \; e^{ -  \frac{i}{r}  } \, , \; i=1,2,... \, ; \;\; r >> 1 \, ,
\end{eqnarray}
where $n_{i}^{*}$, $i=1,2,...,g$, are the coordinates of the most probable occupation vector  $\vec n$ (\ref{rozklad n}).
The above equation gives the probability that a randomly selected worker is in the i$th$ sector,  provided that  the economy is in the state ${\vec n}^{*}$ $= (n_{1}^{*},n_{2}^{*},$ $...,n_{g}^{*})$.

Therefore, to compare both methods the integration of the obtained probability distribution (\ref{rozklad koncowy A}) in the segments $(a_{i},a_{i+1})$  is indispensable.
Firstly, let us consider the case for which the "width" of the sectors is equal to $a_{0} = a_{min}$, the smallest (absolute) level of the productivity. With this assumption $a_{i} = i \, a_{min}$, where $a_{min}=a_{0} > 0$, (\ref{ciag a i}),  the result is obtained:
\begin{eqnarray}
\label{rozklad koncowy A dysktretny}
P\left( i \right) = \int_{i a_{0}}^{(i+1) a_{0}} \! da \; p\left( a \right) = \left( 1 - e^{-1/(r - 1 )} \right) e^{- (i - 1)/(r-1) }
 \quad \;  {\rm for}  \; i=1,2,... \; ,
\end{eqnarray}
which in the limit $r>>1$ gives:
\begin{eqnarray}
\label{rozklad koncowy A dysktretny przybl}
P\left( i \right) \approx  (\frac{1}{r} + \frac{1}{2 r^2}) \, \left(e^{- \frac{i}{r}} \; + \; \frac{1}{r} \right)
 \;\;  {\rm for} \quad \;\; i= 1,2,...\;  \;{ \rm and} \;\; a_{0} > 0  \; , \; r >> 1 \, .
\end{eqnarray}
Secondly, let us consider the case of $a_{0}=0$ with the "width" of the sectors equal to $\delta a$. This leads to
\begin{eqnarray}
\label{rozklad koncowy A dysktretny a0 = 0}
P\left( i \right) = \int_{(i-1) \delta a}^{i \delta a} \! da \; p\left( a \right) = \left( -1 + e^{1/\,\tilde{r}} \, \right) e^{- i/\,\tilde{r} }
 , \; i=1,2,... \;   {\rm for}  \; a_{0}=0 \; ,
\end{eqnarray}
where
\begin{eqnarray}
\label{defin r tilda}
\tilde{r} \equiv \frac{D/n}{\delta a} \; ,
\end{eqnarray}
has been introduced
instead of $r$ (\ref{defin r}). In this case for  $\tilde{r} >> 1$ one can finally get:
\begin{eqnarray}
\label{rozklad koncowy A dysktretny a0 = 0 oraz r duze}
P\left( i \right) \approx  \left( \frac{1}{\tilde{r}} + \frac{1}{2 \;\tilde{r}^2} \right) \, e^{- i/\,\tilde{r} } \; , \; i=1,2,... \; , \;  {\rm for} \quad \; a_{0}=0 \; ; \; \tilde{r} >> 1 \; .
\end{eqnarray}
This means that for large  $\tilde{r}$ both methods give the same results in the first order approximation. Note that
the solution (\ref{rozklad koncowy A dysktretny a0 = 0}) is exact and in the AYM approach some additional assumptions have to be made to obtain the  final formula.


\section{Conclusions}
\label{Conclusions}

The Aoki-Yoshikawa model, although relatively simple, gives interesting results and can be used as a starting point for various analyses. Here we have adopted methods used in theoretical physics to generalize the model and as such these methods refer to the same
phenomenon as the original one \cite{aoki}  (see also Figure~1 in Section~\ref{The solution of the generating equation} and the discussion in \cite{Scalas-Garibaldi-2}). Our approach allows  for exact solutions. The original AY model and the approach presented in this paper agree in first order approximation. Models of phenomena constructed within the proposed approach  can be used as tests of the Extreme Physical Information Principle and we envisage  successful application of these methods in other fields of research \cite{Frieden-Soffer,frieden,Dziekuje-za-EPR-Bohm}.

The original
EPI method
was invented by Frieden and Soffer \cite{Frieden-Soffer,frieden}. They, together with Plastino and Plastino,
put the solution of the (differential) information principles for various EPI models
into practice \cite{Frieden-Soffer}.
Nevertheless, the derivation of the generating equation (\ref{row generujace dla amplitud w EPR}) for probability distribution of the level of productivity (\ref{rozklad koncowy A})
(which could have been inferred by the comparison with the Boltzmann distribution \cite{frieden}) differs from the one used in the original Frieden-Soffer approach \cite{frieden}.
The main difference in the presented derivation
is that in this paper the observed physical information
used directly in the structural information principle
was consistently obtained from the analyticity condition
of the log-likelihood function \cite{Dziekuje informacja_2,Dziekuje-za-EPR-Bohm}
without any jump from its
{\it analytic} to its {\it metric form}.
Only then, the generating equation (\ref{row generujace dla amplitud w EPR}) and  (\ref{produkcyjnosc row generujace})
was derived. This allows consecutively to obtain the probability distribution (\ref{rozklad koncowy A}) of the level of productivity for the statistical informational generalization of the Aoki-Yoshikawa sectoral productivity model.
%

\section*{Acknowledgments}
The work has been supported by the project {\bf Quantum games: theory and implementations} financed by the {\bf National Science Center} under the contract no {\bf DEC-2011/01/B/ST6/07197}.
The contribution of J. Syska to the work has also been supported by the Modelling Research Institute, 40-059 Katowice, Drzyma{\l}y 7/5, Poland.

%

\end{document}